# VELOCITY DISPERSIONS AND X-RAY TEMPERATURES OF GALAXY CLUSTERS

M. GIRARDI, D. FADDA,
G. GIURICIN, F. MARDIROSSIAN, M. MEZZETTI

*Dipartimento di Astronomia, Università degli Studi di Trieste,*
*SISSA, via Beirut 4, 34013 - Trieste, Italy,*
*Email: girardi@tsmi19.sissa.it; fadda@tsmi19.sissa.it;*
*giuricin@tsmi19.sissa.it; mardirossian@tsmi19.sissa.it; mezzetti@tsmi19.sissa.it*

and

A. BIVIANO

*Leiden Sterrewacht, Postbus 9513, Niels Bohrweg 2, 2300 RA Leiden, The Netherlands,*
*Email: biviano@strw.LeidenUniv.nl*

## ABSTRACT

Using a large and well-controlled sample of clusters of galaxies, we investigate the relation between cluster velocity dispersions and X-ray temperatures of intra-cluster gas.

The cluster selection is based on nonparametric methods. In particular, we present the 2-D optical maps of our sample clusters, obtained via the kernel adaptive technique, using an optimized smoothing parameter.

In order to obtain a reliable estimate of the total velocity dispersion of a cluster, independent of the level of anisotropies in galaxy orbits, we analyze the integrated velocity dispersion profiles over increasing distances from the cluster centers. Both increasing and decreasing integrated profiles are found, but the general trend is a flattening of the integrated velocity dispersion profile at the largest radii, thus enabling us to take the asymptotic value of the integrated profile as an estimate of the total velocity dispersion which is independent of possible anisotropies.

Distortions in the velocity fields, the effect of close clusters, the presence of substructures, and the presence of a population of (spiral) galaxies not in virial equilibrium with the cluster potential, are taken into account for reducing the errors in the estimate of the cluster velocity dispersions.

Using our final sample of 37 clusters, for which a reliable estimate of the velocity dispersion could be obtained, we derive a relation between the velocity dispersions and the X-ray temperatures, with a scatter reduced by more than 30% with respect to previous works.

A $\chi^2$ fit to the temperature-velocity dispersion relation does not exclude the hypothesis that the ratio between galaxy and gas energy density (the so-called $\beta_{spec}$) is a constant for all clusters. In particular, the value of $\beta_{spec} = 1$, corresponding to energy equipartition, is acceptable.

However, the large data scatter in $\sigma$-$T$ relation may suggest the presence of intrinsic dispersion. This intrinsic dispersion may be due to spurious effects (we consider the effect of cluster ellipticity), as well as to physical reasons, different values of $\beta_{spec}$ pertaining to clusters with different properties.

*Subject headings:* galaxies: clusters of – X-rays: galaxies

## 1. INTRODUCTION

The comparison between X-ray and optical properties of galaxy clusters can help us to understand the structure, dynamics, and evolution of these galaxy systems. The availability in the literature of large samples of X-ray global quantities (see, e.g., David et al. 1993), as well as optical quantities for cluster galaxies (see, e.g., Girardi et al. 1993), allows an accurate determination of the relations existing between X-ray and optical quantities.

Particularly interesting is the relation between the (line of sight) dispersion in the velocity distribution of cluster galaxies (hereafter $\sigma$) and the X-ray temperature of ICM (hereafter $T$), since both quantities are connected to the gravitational potential of the cluster. This relation is expected to be strong only if the gas and the galaxies are in dynamical equilibrium with the cluster potential. In this case, if the thermal conduction is efficient, the X-ray temperature does not depend on distance from cluster center, and so its value measured in the central region is linked to the total gravitational potential. Moreover, the velocity dispersion measured on a galaxy population tracing the whole cluster (hereafter total $\sigma$) is directly linked to the total gravitational potential via the virial theorem. This implies also that the total kinetic energy of galaxies is independent of velocity asymmetries (The & White 1986; Merritt 1988).

A related issue is the determination of the ratio between the specific energies of the galaxies and the gas $\beta_{spec} = \frac{\sigma^2}{kT/\mu m_p}$, where $\mu$ is the mean molecular weight and $m_p$ the proton mass (see, e.g., Sarazin, 1986).

These topics have already been addressed by many authors (see, e.g., Smith, Mushotzky, & Serlemitsos 1979; Mushotzky 1984; Mushotzky 1988; Evrard 1990; Edge & Stewart 1991a; Lubin & Bahcall 1993), who always found a very large scatter both in the $\sigma$-$T$ relation, and in the mean value of $\beta_{spec}$.

This scatter can arise from a deviation of the observed temperature from the "virial" one, i.e. that expected if the gas is in dynamical equilibrium with the cluster potential. For instance, the gas may be incompletely thermalized, because of an ongoing process of cluster formation through the merger of group sub-units, as suggested by observations (see, e.g., Zabludoff & Zaritsky 1995) and by numerical simulations (see, e.g., Evrard 1990). Hot gas injection from galaxies by supernova explosions (see, e.g., Sarazin 1986; White 1991) can raise the observed temperature over the virial one; if part of the support against gravitation is provided by turbulence and magnetic fields (see, e.g., Miralda-Escudé & Babul 1994; Loeb & Mao 1994; Wu 1994), the observed temperature will be lower than the virial one. Moreover temperature gradients may indicate that thermal conduction is not so efficient. A preliminary analysis of ASCA data indicates that, with the exception of central regions, where cooling flows are often present, the clusters are isothermal to within 15% out to 0.5 $h^{-1} Mpc$ (Mushotzky 1994). This may not be a general property of all clusters, however (see, e.g., Eyles et al. 1991; Briel & Henry 1994), and nothing is known about the temperatures of the more external cluster regions.

Not only $T$, but also $\sigma$ is probably a source of scatter in the $T$–$\sigma$ relation. Although the presence of anisotropies in the galaxy orbits does not affect the value of the total spatial (or projected) $\sigma$ (The & White 1986; Merritt 1988), it can strongly influence $\sigma$ as computed on the central cluster region. The velocity dispersion profiles (VDP in the following) are different for different clusters and still poorly known on average (see, e.g., Kent & Sargent 1983; Merritt 1987; Sharples, Ellis, & Gray 1988), so that a measure of $\sigma$ at a given distance from the cluster center is a poor predictor of the total $\sigma$ in the cluster.

Cluster asphericity may pose another problem since the observed projected $\sigma$ is related to the spatial $\sigma$ via a projection factor depending on cluster ellipticity and inclination. The 2-D galaxy distributions and X-ray maps show that several clusters are elongated (Plionis, Barrow, & Frenk 1991; Jones & Forman 1992; Struble & Ftaclas 1994).

There is evidence that $\sigma$, as computed on the spiral galaxy population, is higher than $\sigma$ as computed on the population of ellipticals, thus suggesting that spirals are far from being virialized in the cluster potential (see, e.g., Sodré et al. 1989; Biviano et al. 1992; Scodeggio et al. 1994).

More generally, we cannot expect $\sigma$ and $T$ to be good indicators of the cluster potential when the cluster is still far from dynamical equilibrium, as shown, e.g., by the presence of subclustering both in the galaxy and in the X-ray emitting gas distributions (see, e.g., Fitchett 1988; Escalera et al. 1994; Slezak, Durret, & Gerbal 1994; West 1994). Although the existence of cluster subclustering is well established, it is not yet well understood how much it influences cluster dynamics. In particular, some authors (e.g. Fitchett & Webster 1987; Edge & Stewart 1991a) have claimed that the presence of substructures may bias the estimate of velocity dispersion. Recent X-ray data show that both clusters A2256 and A754 have regions with different $T$ providing evidence for an ongoing merging event (Briel & Henry 1994; Henry & Briel 1995; Zabludoff & Zaritsky 1995). Numerical simulations show that the galaxy velocity dispersion and the gas temperature both increase when (sub)clusters collide (e.g. Schindler & Boehringer 1993; Schindler and Mueller 1993; Roettinger, Burns, & Loken 1993). However, further numerical simulations are required to examine simultaneous variations of $\sigma$ and $T$.

In this paper we re-examine the relation between $\sigma$ and $T$ by taking into account the presence of velocity anisotropies, the possible tidal effects due to close clusters, the presence of substructures, the existence of velocity gradients in cluster fields, the kinematical differences between spiral and elliptical populations, and the effect of cluster asphericity. The main purpose of this paper is to make a reliable estimate of the total velocity dispersion of a cluster, independent of the level of anisotropies in galaxy orbits, through an analysis of the integrated profiles of velocity dispersion over increasing distances from the cluster center.

In § 2 we describe our data-sample; in § 3 we present our results and provide the relevant discussion; in § 4 we present a summary and draw our conclusions. Throughout, all errors are at the 68% confidence level (hereafter c.l.), while the Hubble constant is 100 $h^{-1}$ km s$^{-1}$ $Mpc^{-1}$.

## 2. THE DATA SAMPLE

Our analysis is based on a sample of 43 clusters, each sampled at least up to half an Abell radius, with at least 25 cluster members having available redshifts within 1 $h^{-1}$ $Mpc$ from the center (membership criteria for cluster galaxies are defined in the next section), and with available X-ray temperatures of the intracluster gas (David et al. 1993). These selection criteria were chosen in order to allow the determination of the total $\sigma$ in each cluster through analysis of the VDP. In order to achieve a sufficiently homogenous sample, the galaxy redshifts in each cluster have usually been taken from one reference source only, or from different sources, only when redshifts from these different sources proved to be compatible.

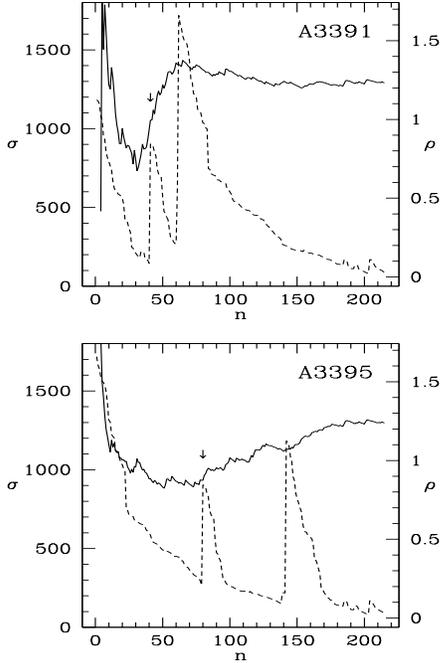

Figure 2: Velocity dispersion (solid line) and galaxy density (dashed line) along the sequences beginning with the peak of the clusters A3391 (upper figure) and A3395 (lower figure). Arrows indicate the point where the velocity dispersion is taken.

We applied homogenous procedures to the study of the optical data of these 43 clusters; we used robust mean and scale estimates (computed with the ROSTAT routines kindly provided by T. Beers – see Beers, Flynn, & Gebhardt 1990). In Table 1 we list the cluster names (Col. 1), the references to the redshift data (Col. 2), the X-ray temperatures and their errors (Col. 3), as given by David et al. (1993), and the references to the adopted X-ray centers (Col. 4). The upper limit in the observed value of $T$ for the cluster A1142 is not available; we took it to be equal to the observed $T$ plus the largest error on $T$ in our sample.

In order to consider the possible influence of neighbouring clusters, for each object we looked for the presence of ACO clusters (Abell *et al.* 1989) in a region of $\sim 2\ h^{-1}\ Mpc$ from the cluster center and in a redshift range of $\sim 0.01$. We identified as neighbours the cluster pairs A399/A401, A3391/A3395. We considered together the two clusters of each pair in the initial cluster selection in the redshift space. We used the same treatment also for the cluster pairs A2063/MKW3S, A2634/A2666, A3558/SC1329-314 presented together by the authors. The influence of close clusters on VDPs is considered in § 3.3.

### 2.1. CLUSTER MEMBERSHIP

In order to assign cluster membership, we have used both the velocity and the spatial distribution of galaxies projected into the cluster area. We have analyzed each cluster velocity distribution via the adaptive kernel technique (see Appendix A), that provides the significance of each detected peak in the velocity distribution, as well as an estimate of the extent of the overlapping between contiguous peaks. A peak is considered to be real if it is detected at a c.l. ≥ 99%; the main cluster body is then naturally identified as the highest significant peak. All galaxies not belonging to this peak are rejected as non-cluster members. We stress that this procedure is non-parametric, like the method of weighted gaps which is frequently used in the literature (see, e.g., Beers, Flynn, & Gebhardt 1990; Beers et al. 1991; Girardi et al. 1993); the advantage of the adaptive kernel technique vs. the weighted-gap one is that the former allows us to represent the data without binning and to quantify the overlapping between contiguous peaks.

Table 1: The Initial Cluster Sample. Velocity and X-ray center references at the end of the paper (Notes to the Tables)

| Name (1) | Vel. Refs. (2) | $T\ (keV)$ (3) | X-ray center Refs. (4) |
|---|---|---|---|
| A 85 | [1,2] | $6.2^{+0.2}_{-0.3}$ | [44] |
| A 119 | [3] | $5.9^{+0.6}_{-0.6}$ | [45] |
| A 193 | [4] | $4.2^{+1.0}_{-0.5}$ | [46] |
| A 262 | [5,6,7] | $2.4^{+0.2}_{-0.1}$ | [44] |
| A 399 | [4] | $5.8^{+0.8}_{-0.7}$ | [45] |
| A 400 | [8] | $2.5^{+0.4}_{-0.4}$ | [45] |
| A 401 | [4] | $7.8^{+0.6}_{-0.6}$ | [45] |
| A 426 Perseus | [9] | $6.3^{+0.2}_{-0.2}$ | [44] |
| A 496 | [10] | $3.9^{+0.1}_{-0.1}$ | [45] |
| A 539 | [11] | $3.0^{+0.5}_{-0.4}$ | [45] |
| A 548S | [12] | $2.4^{+0.7}_{-0.5}$ | [47] |
| A 576 | [13] | $4.3^{+0.3}_{-0.3}$ | [46] |
| A 754 | [12] | $9.1^{+0.7}_{-0.6}$ | [45] |
| A1060 Hydra | [14,15] | $3.9^{+0.2}_{-0.2}$ | [44] |
| A1142 | [16] | $3.7^{+5.4}_{-2.0}$ | [48] |
| A1367 | [6,17,18,19] | $3.7^{+0.2}_{-0.1}$ | [47] |
| A1644 | [15] | $4.7^{+0.5}_{-0.5}$ | [44] |
| A1656 Coma | [20] | $8.3^{+0.6}_{-0.5}$ | [45] |
| A1689 | [21] | $10.1^{+5.4}_{-2.8}$ | [44] |
| A1736 | [12] | $4.6^{+0.7}_{-0.6}$ | [49] |
| A1795 | [4] | $5.8^{+0.3}_{-0.2}$ | [46] |
| A2052 | [1,22] | $3.1^{+0.2}_{-0.2}$ | [45] |
| A2063 | [2,4] | $4.1^{+0.6}_{-0.6}$ | [48] |
| A2107 | [23] | $4.2^{+1.9}_{-1.1}$ | [50] |
| A2151 Hercules | [12] | $3.8^{+0.7}_{-0.5}$ | [51] |
| A2199 | [24] | $4.5^{+0.2}_{-0.1}$ | [51] |
| A2256 | [25] | $7.3^{+0.5}_{-0.4}$ | [52] |
| A2593 | [2] | $3.1^{+1.5}_{-0.9}$ | [44] |
| A2634 | [26,27,28] [29,30,31] | $3.4^{+0.2}_{-0.3}$ | [31] |
| A2670 | [32] | $3.9^{+1.6}_{-0.8}$ | [48] |
| A2877 | [1] | $3.5^{+1.1}_{-0.8}$ | [47] |
| A3158 | [33,34] | $5.5^{+0.3}_{-0.3}$ | [44] |
| A3266 | [21] | $6.2^{+0.5}_{-0.4}$ | [47] |
| A3391 | [21] | $5.2^{+1.3}_{-0.9}$ | [47] |
| A3395 | [21] | $4.7^{+1.1}_{-0.7}$ | [47] |
| A3526 Centaurus | [35,36] | $3.9^{+0.1}_{-0.1}$ | [44] |
| A3558 Shapley 8 | [14,21,37,38] | $3.8^{+1.2}_{-0.6}$ | [49] |
| A3571 | [39] | $7.6^{+0.7}_{-0.6}$ | [53] |
| A3667 | [40] | $6.5^{+1.0}_{-1.0}$ | [54] |
| S0805 | [1] | $1.4^{+0.3}_{-0.3}$ | [47] |
| MKW3S | [2,4] | $3.0^{+0.3}_{-0.3}$ | [44] |
| MKW4 | [41,42] | $1.7^{+1.7}_{-0.7}$ | [44] |
| VIRGO | [43] | $2.4^{+0.2}_{-0.2}$ | [46] |

Table 2: Effects of Galaxy Morphology. References are at the end of the paper (Notes to the Tables)

| Name (1) | $N_e$ (2) | $N_l$ (3) | $R^*$ (4) | $P_m$ (5) | $P_F$ (6) | Ref. (7) |
|---|---|---|---|---|---|---|
| A 119 | 58 | 13 | 0.98 | 0.83 | 0.82 | [1] |
| A 400 | 58 | 24 | 1.24 | 0.98 | 0.57 | [2,3,4,5] |
| A 426 Perseus | 54 | 45 | 2.69 | 0.69 | 0.36 | [4,5,6,7] |
| A 496 | 54 | 18 | 0.97 | 0.56 | 0.34 | [3] |
| A 539 | 46 | 35 | 4.76 | 0.92 | 0.94 | [4,5,8] |
| A 754 | 62 | 14 | 1.90 | 0.06 | 0.36 | [9] |
| A1060 Hydra | 82 | 56 | 2.01 | >0.99 | 0.25 | [10,11] |
| A1644 | 70 | 20 | 1.77 | 0.95 | 0.51 | [9] |
| A1656 Coma | 173 | 85 | 4.93 | 0.25 | 0.98 | [12] |
| A2063 | 37 | 17 | 1.37 | 0.42 | 0.78 | [3] |
| A2151 Hercules | 56 | 44 | 1.45 | 0.99 | 0.10 | [9] |
| A2256 | 46 | 13 | 0.92 | 0.49 | 0.98 | [3] |
| A2634 | 61 | 22 | 0.69 | 0.44 | >0.99 | [3] |
| A2670 | 90 | 26 | 1.12 | 0.60 | 0.89 | [13] |
| A2877 | 35 | 10 | 0.96 | 0.88 | >0.99 | [3] |
| S0805 | 74 | 39 | 1.77 | 0.62 | 0.48 | [14] |
| VIRGO | 179 | 246 | 2.06 | 0.32 | >0.99 | [15] |

Table 3: The final sample. The asterisk indicates that only early-type galaxies are used in the analysis.

| Name (1) | $N$ (2) | $R$ (3) | $RGD$ (4) | $P_{DS}$ (5) |
|---|---|---|---|---|
| A 85 | 131 | 4.49 | 0.75 | 0.359 |
| A 119 | 80 | 1.46 | 0.33 | 0.987 |
| A 193 | 58 | 0.89 | - | 0.651 |
| A 262 | 78 | 4.95 | 0.37 | 0.602 |
| A 399 | 92 | 1.71 | - | 0.285 |
| A 400* | 59 | 1.22 | 0.23 | 0.391 |
| A 401 | 123 | 2.11 | 0.45 | 0.358 |
| A 426 Perseus | 127 | 2.70 | 0.27 | 0.144 |
| A 496 | 151 | 1.38 | 0.47 | 0.978 |
| A 539 | 102 | 4.87 | 0.07 | 0.998 |
| A 576 | 48 | 1.22 | - | 0.059 |
| A 754 | 83 | 2.47 | >0.99 | 0.458 |
| A1060* Hydra | 82 | 2.08 | 0.16 | 0.894 |
| A1142 | 44 | 2.21 | >0.99 | 0.618 |
| A1644 | 92 | 1.95 | - | 0.895 |
| A1656* Coma | 170 | 4.98 | 0.93 | 0.296 |
| A1795 | 87 | 1.76 | 0.12 | 0.346 |
| A2052 | 62 | 1.37 | - | 0.593 |
| A2063 | 91 | 4.77 | 0.05 | 0.802 |
| A2107 | 68 | 0.99 | - | 0.801 |
| A2151* Hercules | 58 | 1.88 | 0.41 | 0.980 |
| A2199 | 51 | 3.22 | 0.28 | 0.292 |
| A2256* | 47 | 0.93 | - | 0.885 |
| A2593 | 37 | 1.39 | 0.36 | 0.534 |
| A2634* | 69 | 0.87 | 0.56 | 0.335 |
| A2670 | 215 | 2.24 | 0.15 | 0.920 |
| A2877* | 37 | 1.04 | - | 0.662 |
| A3158 | 35 | 1.44 | - | 0.629 |
| A3266 | 132 | 1.11 | 0.55 | 0.479 |
| A3391 | 55 | 0.90 | 0.49 | 0.271 |
| A3395 | 107 | 1.14 | 0.34 | 0.982 |
| A3558 Shapley 8 | 206 | 0.98 | 0.49 | 0.988 |
| A3571 | 70 | 0.96 | - | 0.057 |
| A3667 | 123 | 2.25 | 0.37 | 0.697 |
| S0805 | 120 | 1.90 | 0.72 | 0.558 |
| MKW3S | 30 | 3.47 | - | 0.802 |
| MKW4 | 53 | 4.88 | 0.99 | 0.964 |
| VIRGO* | 180 | 2.03 | 0.66 | 0.996 |

We suspect that clusters with strong (> 20%) peak overlapping in their velocity distributions, viz. A548(60%), A1367(66%), A1689(20%), A1736(24%), A3526(56%), are very far from dynamical equilibrium and we discuss them separately (see § 3.4). The following analysis concerns the remaining 38 clusters.

We applied to each cluster sample a 2-D adaptive kernel analysis (see Appendix A). Using the 2-D adaptive kernel technique, we obtained, as in the 1-D case, the significance, the number of galaxies and the galaxy density of each peak.

In order to separate the clusters from their neighbours we used the 2-D analysis, choosing for each cluster the corresponding peak (A399, A401, A3391, A3395). For clusters A2634 and A3558 we considered all the galaxies within 2 $h^{-1}$ $Mpc$ and 1 $h^{-1}$ $Mpc$ from the cluster center in order to separate them from their neighbours A2666 and SC1329-314, respectively. Moreover, we never considered galaxies beyond 5 $h^{-1}$ $Mpc$ from their cluster center.

Since the velocity dispersion depends on galaxy velocities, but also on the distribution of galaxies in space, it would be best to have samples complete to a limiting magnitude. In particular, if a deeper sampling of galaxies (in terms of magnitudes) has been made by observers in the central regions, the velocity dispersion of the galaxies in the central region will weigh too much in the computation of the total velocity dispersion, based on the total galaxy sample. This is very dangerous if the VDP is rapidly increasing/decreasing. Therefore, for all our clusters with available galaxy magnitudes, we looked at the galaxy magnitudes as a function of clustercentric distance.

We have noted that some clusters are sampled more deeply close to the center than they are outside. We have rejected the fainter galaxies of these clusters so as to eliminate any trend of the limiting magnitude to vary with clustercentric distance. Incompleteness effects in the external regions of some clusters are not very important, since the integral VDP flattens out in these regions.

2.2. EFFECTS OF GALAXY MORPHOLOGY

In 17 (out of 38) clusters we have at least partial information on galaxy morphologies. In order to test for different means and variances in velocity distributions of early- and late-type galaxies, we applied the standard means-test and F-test (Press et al. 1992), to the subsamples of early and late-type galaxies, within the largest area occupied by both galaxy populations.

Table 2 lists the 17 clusters used in the morphological analysis. In Col. (1) we list the cluster names; in Cols. (2) and (3) the numbers of early-type $N_e$ and late-type $N_l$ galaxies (respectively), within the radius $R^*$ listed in Col. (4); in Cols. (5) and (6) $P_m$ and $P_F$, the probabilities that means and velocity dis-persions of early and late-type galaxy velocity distributions are different, according to the means-test and F-test, respectively; in Col. (7) the relevant morphology reference sources.

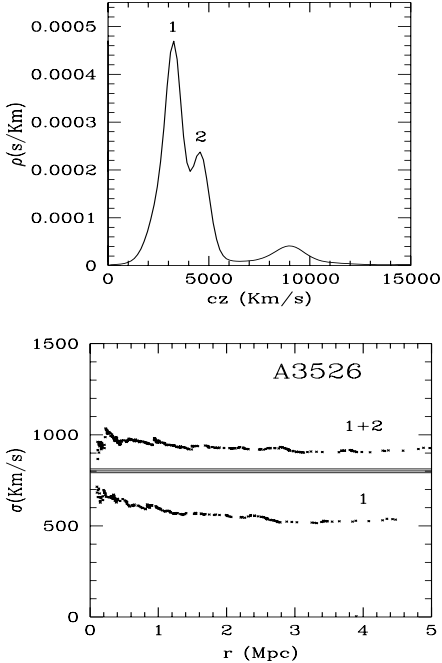

Figure 3: In the upper figure we show the double-peak velocity distribution of cluster A3526. In the lower figure we plot the VDP corresponding to the most important peak and the VDP, with higher $\sigma$, corresponding to the two peaks together.

When the probability for the early- and late-type galaxy populations to have different velocity distributions was larger than 0.95, according to at least one of the two tests, we chose to consider only early-type galaxies (this was the case in the clusters: A400, A1060, A2151, and A1656, A2256, A2634, A2877 and Virgo, according to the means- and F-test, respectively). The selection procedure (described in § 2.1) was then repeated on these nine clusters, by considering only the early-type galaxy population; however, we found no substantial difference with respect to an *a posteriori* selection.

### 2.3. THE FINAL SAMPLE

Our final cluster sample contains 38 clusters. Fig. 1 represents the 2-D galaxy distributions as adaptive kernel density contour maps (sometimes zoomed in on a central region). Moreover, in Fig. 1 we indicate the position of the X-ray center as found in the literature (see Tab. 1 and the discussion in § 3.2) and we plot a 10-arcminutes radius circle to give a rough indication of the region used in the $T$ estimates (only in the case of Virgo do we take a radius of 40 arcminutes). We applied the 2-D adaptive kernel method to this final sample and we estimated the relative importance of peaks in each cluster by normalizing the galaxy density of the peak to the density of the most significant peak. We also computed the probability of substructure $P_{DS}$ according to the test of Dressler & Schectman (1988a). The final sample is presented in Table 3. In Col. (1) we list the cluster names; in Cols. (2) and (3) the number of galaxy members $N$, and the cluster extension $R$ (in Mpc) of the final sample, respectively; in Col. (4) the relative galaxy density, $RGD$, of the secondary peak, when statistically significant; and in Col. (5) the $P_{DS}$ substructure probability.

## 3. RESULTS AND DISCUSSION

### 3.1. VELOCITY GRADIENTS IN THE CLUSTER VELOCITY FIELD

The cluster velocity field may be influenced by the existence of other structures on larger scales such as a nearby cluster or the supercluster to which the cluster belongs, or filaments, and so on. Each asymmetrical effect could produce a velocity gradient in the cluster velocity field. We analyzed the presence of velocity gradients, performing for each cluster a multiple linear regression fit to the observed velocities with respect to the galaxy positions (ascension and declination) in the plane of the sky (see, e.g., den Hartog & Katgert 1994). For each cluster we computed the velocity gradients and the coefficient of multiple determination $R^2$, $0 \leq R^2 \leq 1$, which measures the amount of deviation of the dependent variable due to the set of the two independent variables (e.g. NAG Fortran Workstation Handbook, 1991). We tested the significance of the fitted velocity gradients using 1000 Monte-Carlo simulations for each cluster, performed by randomly shuffling the galaxy velocities and computing the $R^2$ coefficient every time. For each cluster we defined the significance of velocity gradients as the fraction of times in which the $R^2$ we obtain from simulations is smaller than the observed $R^2$. This significance is greater than 99% for five clusters: A399, A401, A2107, S805, and Virgo. For these five clusters we applied a correction by subtracting the velocity gradients from each galaxy velocity and renormalizing the velocities so as to leave their average unchanged. However, the correction has little effect both on the shape of VDP (discussed in § 3.2) and on the total velocity dispersion (the mean absolute correction is $\sim -30$ km s$^{-1}$).

### 3.2. VELOCITY DISPERSION PROFILES

We considered the line of sight velocity dispersions integrated on larger and larger radii. It is important to analyse the integral VDP, because of the particular significance of the total value of velocity dispersion (which is independent of velocity anisotropies, see § 1) as well as for establishing the dependence of the estimate of $\sigma$ on the sample extension in each cluster. In Fig. 1 we plotted the integral VDP vs. the clustercentric distance for each cluster. The velocity dispersions are computed using the robust velocity dispersion (e.g. Beers, Flynn, and Gebhardt 1990); the bands in Fig. 1 represent the bootstrap errors (at 68% c.l.). The horizontal lines show the values of the velocity dispersion, with their respective error bands, obtained from the temperatures listed in Tab. 1 under the condition of perfect galaxy/gas energy equipartition, i.e. $\beta_{spec} = 1$, and molecular weight $\mu = 0.58$, which is the adopted average value for clusters (Edge & Stewart 1991a).

The crucial importance of the choice of cluster center and the possible differences among centers determined by different procedures are pointed out in several papers (e.g. Beers & Tonry 1985; Rhee & Latour 1989). Since we want to compare our velocity dispersions with X-ray temperatures, we chose the X-ray centers published in the literature. However, the computed VDPs at large radii are not strongly affected by different choices of cluster center (we used the position of the first-ranked galaxy, and the peak of density galaxy distribution, too). The contour maps in Fig. 1 show that the X-ray and galaxy density centers are similar. In several clusters (e.g. Virgo) this agreement is better when the early-type rather than the late-type galaxy population is considered.

### 3.3. TOTAL VELOCITY DISPERSIONS

The trend of integrated VDP happens to be different for different clusters in the central region, but it is generally flat in the external region. The increasing/decreasing trends of VDP in the central region may be due both to velocity anisotropies and to dark matter distribution, but it is not easy to disentangle these effects (e.g. Merritt 1988) and we postpone this analysis to a future work. However, the flatness of VDPs in the external region suggests that, there, possible velocity anisotropies no longer affect the value of $\sigma$.

Moreover, the VDP flattens out at about 1 $h^{-1}\,Mpc$. For cluster masses of about $10^{14} - 10^{15}\,h^{-1}\,M_\odot$, this size correponds to that of the collapsed and virialized region. So we expect that the asymptotic value of $\sigma$ is representative of the total kinetic energy of galaxies.

For each cluster we take the final value of $\sigma$ in the observed VDPs as a fair estimate of the total $\sigma$. There are, however, a few exceptions. Clusters A3391 and A3395, which close to one another, both have a VDP strongly increasing towards the external regions, probably because of the presence of a neighbouring cluster. For these close clusters, we computed $\sigma$ along a sequence of galaxies with decreasing density starting from a density maximum (see Appendix A). In Fig 2 we plot the velocity dispersion (as well the galaxy density) along the sequences beginning with the peak of the clusters A3391 and A3395. In the A3391/A3395 region there are only 3 significant density peaks: two correspond to the clusters and the third is an intermediate group (see also the map in Fig. 1). For both A3391 and A3395 we adopted the value of $\sigma$ obtained before we encounter the following density peak, representing the intermediate group (see Fig. 2a,b). These values of $\sigma$ are similar to the values of $\sigma$ in the lowest point of VDP (see Fig 1).

The adopted $\sigma$ values are presented in Table 4. In Col. (1) we list cluster names; in Cols. (2) and (3) $\sigma$ and $\beta_{spec} = \frac{\sigma^2}{kT/\mu m_p}$, with their respective errors.

### 3.4. THE EFFECTS OF SUBSTRUCTURES

Up to now a large fraction of observed clusters have shown substructures, whose detection increases with the growing availability of the data (see the Coma cluster, e.g. Escalera et al. 1993). So the question is not whether a cluster has or not has substructures, but rather how far the cluster is from dynamical equilibrium. As described in §2.1, we decided to exclude from our analysis 5 clusters with strong overlapping of the peaks in the velocity distribution. As an example, in Fig. 3a we show the double-peak velocity distribution of A3526 (see Lucey, Currie & Dickens 1986). In Fig. 3b we plot the VDP corresponding to the most important peak and the VDP, with higher $\sigma$, corresponding to the two peaks taken together. We think that these clusters are probably very far from dynamical equilibrium, and perhaps they are examples of merging clusters. The same may be true for other clusters, however; for example, A754 is apparently bimodal in the 2-D map (see also Zabludoff & Zaritsky 1995). We shall discuss it in the next section.

Less apparent substructures are debatable, and we are far from understanding their influence on the overall cluster dynamics (e.g. González-Casado, Mamon, & Salvador-Solé 1994). Moreover, different methods of analysis are suitable for detect-ing different kinds of substructure (see, e.g., West, Oemler, & Dekler 1988). As the question remains open, in this paper we preferred not to apply any corrections to account for the presence of cluster substructures. Bird, Mushotzky, & Metzler (1995) corrected for the presence of substructures by using the KMM method of decomposition in a set of gaussian velocity distributions (see also Ashman et al. 1994). It is clear, however, that this correction cannot take into account the presence of velocity anisotropies which produce non-gaussian velocity distributions. Our non-parametric selection of clusters takes this (possible) problem into account.

Table 4: The velocity dispersion and $\beta_{spec}$. The asterisk indicates that only early-type galaxies are used in the analysis.

| Name (1) | $\sigma\,(Km/s)$ (2) | $\beta_{spec}$ (3) |
|---|---|---|
| A 85 | $1069._{-92.}^{+105.}$ | $1.12_{-0.25}^{+0.26}$ |
| A 119 | $850._{-92.}^{+108.}$ | $0.74_{-0.24}^{+0.26}$ |
| A 193 | $756._{-83.}^{+119.}$ | $0.82_{-0.28}^{+0.46}$ |
| A 262 | $575._{-43.}^{+69.}$ | $0.83_{-0.16}^{+0.27}$ |
| A 399 | $1195._{-79.}^{+94.}$ | $1.49_{-0.38}^{+0.44}$ |
| A 400* | $607._{-72.}^{+76.}$ | $0.89_{-0.35}^{+0.37}$ |
| A 401 | $1142._{-70.}^{+80.}$ | $1.01_{-0.20}^{+0.22}$ |
| A 426 Perseus | $1284._{-95.}^{+140.}$ | $1.59_{-0.29}^{+0.40}$ |
| A 496 | $750._{-56.}^{+61.}$ | $0.87_{-0.15}^{+0.16}$ |
| A 539 | $747._{-93.}^{+105.}$ | $1.13_{-0.43}^{+0.50}$ |
| A 576 | $1006._{-91.}^{+138.}$ | $1.43_{-0.36}^{+0.49}$ |
| A 754 | $784._{-85.}^{+90.}$ | - |
| A1060* Hydra | $614._{-43.}^{+52.}$ | $0.59_{-0.11}^{+0.13}$ |
| A1142 | $631._{-114.}^{+156.}$ | $0.65_{-0.59}^{+1.27}$ |
| A1644 | $937._{-77.}^{+107.}$ | $1.13_{-0.31}^{+0.38}$ |
| A1656* Coma | $913._{-63.}^{+73.}$ | $0.61_{-0.12}^{+0.14}$ |
| A1795 | $887._{-83.}^{+116.}$ | $0.82_{-0.18}^{+0.26}$ |
| A2052 | $679._{-59.}^{+97.}$ | $0.90_{-0.21}^{+0.32}$ |
| A2063 | $664._{-45.}^{+50.}$ | $0.65_{-0.18}^{+0.19}$ |
| A2107 | $625._{-58.}^{+75.}$ | $0.56_{-0.25}^{+0.39}$ |
| A2151* Hercules | $743._{-68.}^{+95.}$ | $0.88_{-0.28}^{+0.39}$ |
| A2199 | $860._{-83.}^{+134.}$ | $1.00_{-0.21}^{+0.35}$ |
| A2256* | $1279._{-117.}^{+136.}$ | $1.36_{-0.32}^{+0.38}$ |
| A2593 | $700._{-69.}^{+116.}$ | $0.96_{-0.47}^{+0.78}$ |
| A2634* | $705._{-61.}^{+97.}$ | $0.89_{-0.23}^{+0.30}$ |
| A2670 | $983._{-54.}^{+72.}$ | $1.50_{-0.51}^{+0.84}$ |
| A2877* | $748._{-81.}^{+126.}$ | $0.97_{-0.43}^{+0.63}$ |
| A3158 | $1046._{-99.}^{+174.}$ | $1.21_{-0.32}^{+0.47}$ |
| A3266 | $1182._{-85.}^{+100.}$ | $1.37_{-0.28}^{+0.34}$ |
| A3391 | $990._{-128.}^{+254.}$ | $1.14_{-0.49}^{+0.87}$ |
| A3395 | $934._{-100.}^{+123.}$ | $1.12_{-0.41}^{+0.56}$ |
| A3558 Shapley 8 | $997._{-51.}^{+61.}$ | $1.59_{-0.41}^{+0.69}$ |
| A3571 | $1085._{-107.}^{+110.}$ | $0.94_{-0.26}^{+0.28}$ |
| A3667 | $1208._{-84.}^{+95.}$ | $1.36_{-0.40}^{+0.42}$ |
| S0805 | $549._{-48.}^{+52.}$ | $1.30_{-0.51}^{+0.53}$ |
| MKW3S | $612._{-52.}^{+69.}$ | $0.76_{-0.20}^{+0.25}$ |
| MKW4 | $539._{-59.}^{+85.}$ | $1.04_{-0.65}^{+1.36}$ |
| VIRGO* | $643._{-30.}^{+41.}$ | $1.04_{-0.18}^{+0.22}$ |

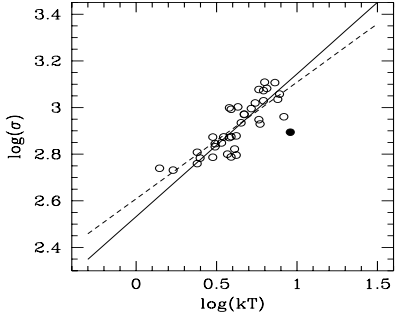

Figure 4: The 38 data point of the final sample. The solid line is the fit on 37 clusters excluding A754 (solid point). The dashed line represents the model with $\beta_{spec} = 1$.

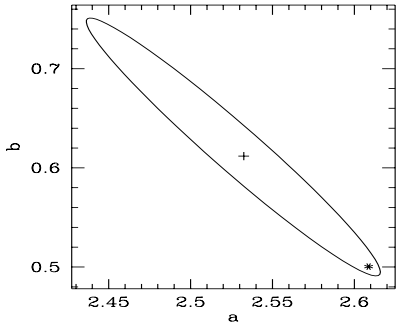

Figure 5: The confidence ellipse at 2 standard deviations (95.4% c.l.), corresponding to $\sigma = 10^a \cdot T^b$ as fitted in eq. 1. The asterisk represents the cluster model of perfect galaxy/gas energy equipartition, i.e. $\beta_{spec} = 1$.

### 3.5. THE $\sigma$ - T RELATION

In order to describe the physical law which links $\sigma$ and $T$, two independent variables, it is necessary to use regression methods which treat the variables symmetrically (see, e.g., Kendall & Stuart 1979). When uncertainties in both variables are significant and known, Isobe et al. (1990) and Feigelson & Babu (1992) recommend a (double) weighted functional regression procedure which accounts for errors in both axes. In this paper we used a Maximum Likelihood estimate of the regression lines (see, e.g., Kendall & Stuart 1979; Press et al. 1992) in order to fit the logarithmic quantities log $\sigma$-log $T$ after having symmetrized the errors of $T$ and $\sigma$ listed in Tabs. 1 and 4, respectively.

We obtained $\sigma = 10^{2.56\pm0.03} \cdot T^{0.56\pm0.05}$. In order to measure the scatter of the data about the line we computed the standard deviation (hereafter s.d.) of the residuals (i.e. the distances between the data and the fitted line). This s.d. is reduced with respect to the one we computed using both data and fitted lines of other authors (Edge & Stewart 1991a; Lubin & Bahcall 1993; and Bird, Mushotzky, & Metzler 1995). Cluster A754, which is also one of the most 2-D structured clusters in our sample, gives the largest residual, at about 3 s.d.'s. Surprisingly the presence of substructure leads either to hotter temperature or to lower velocity dispersion than required by the cluster model with $\beta = 1$. Since A754 is also a well-known substructured cluster (e.g. Fabricant et al. 1986; Slezak, Durret, & Gerbal 1994) and shows evidence of subcluster collision (Zabludoff & Zaritsky 1995) we prefer to exclude A754 from the following analyses.

On the remaining 37 clusters we fitted

$$\sigma = 10^{2.53\pm0.04} \cdot T^{0.61\pm0.05}, \qquad (1)$$

the s.d. of residuals being reduced by more than 30% with respect to previous works (see Tab. 5). In Fig. 4 we plot the data points and the fitted line on 37 clusters. We devoted particular attention to the error analysis. The errors in eq. 1 are projections of the confidence ellipse at the 68% c.l. In order to study the compatibility of a model with the data, in Fig. 5 we plot the confidence ellipse at the 95.4% c.l. (corresponding to 2 s.d.). The model of perfect galaxy/gas energy equipartition, $\beta_{spec} = 1$, represented by an asterisk in Fig. 5, is compatible with our data.

When considering the subsample of 16 clusters with known galaxy morphological information we obtained $\sigma = 10^{2.59\pm0.04} \cdot T^{0.50\pm0.07}$, consistent within the errors with eq. 1.

With the exception of A754, the average value of $\beta_{spec}$ is $1.03 \pm 0.05$.

Edge & Stewart (1991a), having found that the computed $\beta_{spec}$ correlated with $\sigma$ and not with $T$, suggested that the values of $\sigma$ are not reliable. Actually, also in our sample there is a significant correlation between $\beta_{spec}$ and $\sigma$. It is well possible that our values of $\sigma$ are still affected by some uncorrected bias, but in any case it is not possible to draw any conclusion on that by using the correlation between $\beta_{spec}$ and $T$ or $\sigma$. This is due to the fact that $\beta_{spec}$ is *defined* as a function of $T$ and $\sigma$, and this automatically induces correlations between $\beta_{spec}$, $T$ and $\sigma$. This point is extensively described, in a similar context, by Mezzetti et al. (1982). In particular, defining $\rho_{xy}$ as the Pearson's correlation coefficient between $log(x)$ and $log(y)$, and $\Sigma_x$ as the standard deviation of the $log(x)$ variable, it is possible to show that:

$$\rho_{\beta\sigma} = \frac{2\Sigma_\sigma(1 - \rho_{\sigma T}\frac{\Sigma_T}{2\Sigma_\sigma})}{\sqrt{4\Sigma_\sigma^2 + \Sigma_T^2 - 4\rho_{\sigma T}\Sigma_\sigma\Sigma_T}} \qquad (2)$$

and

$$\rho_{\beta T} = \frac{\Sigma_T(2\rho_{\sigma T}\frac{\Sigma_\sigma}{\Sigma_T} - 1)}{\sqrt{4\Sigma_\sigma^2 + \Sigma_T^2 - 4\rho_{\sigma T}\Sigma_\sigma\Sigma_T}}. \qquad (3)$$

So, these two correlations do not contain any further information about the $log(T) - log(\sigma)$ correlation. Moreover, due to the definition of $\beta_{spec}$, this may turn out to be correlated with $log(T)$ and/or $log(\sigma)$ even in the case that $log(T)$ and $log(\sigma)$ are not correlated at all.

If $\beta_{spec}$ is indeed constant for all clusters, the $\sigma$-$T$ relation could be an efficient method for identifying clusters far from dynamical equilibrium when they deviate from the fitted (or expected) relation. In this paper we parametrize the presence of substructure by selecting, from among many possibilities, the 2-D density parameter, $RGD$, and the well known parameter by Dressler & Schectman (1988a), $P_{DS}$. We found no correlation between substructure parameters, $RGD$ and $P_{DS}$, and the (absolute) residuals in our $\sigma$-$T$ relation. However, cluster A754, which has the highest $RGD$ in our sample, is about 3 s.d.'s from the $\sigma$-$T$ relation. It is clear that a deeper understanding of the dynamical effect of substructures is needed before we can reach any definite conclusion.

| Reference | $N_c$ | $\langle \frac{\Delta\sigma}{\sigma} \rangle$ | $\langle \frac{\Delta T}{T} \rangle$ | Fitted Relation | sd of residuals | $\langle \beta_{spec} \rangle$ | $\beta_{spec}$ rms |
|---|---|---|---|---|---|---|---|
| (1) | (2) | (3) | (4) | (5) | (6) | (7) | (8) |
| Edge & Stewart, 1991 | 23 | 0.12 | 0.18 | $\sigma = 10^{2.60\pm0.08} \cdot T^{0.46\pm0.12}$ | 1.46 | 0.91 | 0.38 |
| Lubin & Bahcall, 1993 | 41 | 0.11 | 0.17 | $\sigma = 10^{2.52\pm0.07} \cdot T^{0.60\pm0.11}$ | 1.61 | 1.14 | 0.57 |
| Bird, Mushotzky, & Metzler, 1995 | 22 | 0.17 | 0.21 | $\sigma = 10^{2.50\pm0.09} \cdot T^{0.61\pm0.13}$ | 1.65 | 0.90 | 0.37 |
| This paper | 37 | 0.11 | 0.17 | $\sigma = 10^{2.53\pm0.04} \cdot T^{0.61\pm0.05}$ | 1.00 | 1.03 | 0.29 |

Table 5: Comparison with Other Works. In column 2 the number of clusters in the sample. In columns 3 and 4 the average relative error on $\sigma$ and $T$. In column 6 the s.d. of residuals in $\log\sigma - \log T$ plane normalized to our value (0.054). In columns 7 and 8 the average $\beta_{spec}$, as obtained directly from $\beta_{spec} = \frac{\sigma^2}{kT/\mu m_p}$, and its rms.

### 3.6. THE EFFECT OF ASPHERICITY

According to the $\chi^2$ fit probability, the fit to the data in the $\log T$-$\log\sigma$ plane is acceptable at the 2% c.l.. This result may suggest that the errors do not take into account other sources of scatter, e.g. cluster asphericity.

The effect of cluster asphericity on the estimate of the total velocity dispersion may be estimated directly if one assumes that clusters are axisymmetric (prolate or oblate), that isodensity surfaces are concentric, similar ellipsoids, and that galaxies are distributed in a way similar to the total binding mass. In this case it is possible to evaluate analytically the projection factor between the line-of-sight velocity dispersion and the corresponding 3D value, as a function of the inclination and of the intrinsic shape (eccentricity $\epsilon$ or axial ratio $\beta$) of the cluster (see Appendix B).

Interestingly, even if the projection factor usually differs from the spherically symmetric value (i.e. $\sqrt{3}$), its expectation value, averaged on inclination, differs by less than 3% from $\sqrt{3}$ also in the case $\epsilon = 0.7$, which is the highest observed value in the cluster sample analyzed by Struble and Ftaclas (1994). The same result is obtained if, following Plionis, Barrow & Frenk (1991), one assumes that clusters are prolate and have a gaussian distribution of intrinsic axial ratios with a mean of 0.5 and a s.d. of 0.15. It follows that there is no reason to adopt a projection factor different from $\sqrt{3}$. Adopting the above-mentioned distribution of axial ratios, one obtains that the standard deviation in $\log(\sigma)$ induced by a random orientation of aspherical clusters is 0.04, corresponding to 63% of the observed scatter (see Table 5). Increasing the $\log\sigma$ errors with this value, we improve the $\chi^2$ probability of the fit (30%). Hence it is possible that asphericity is partially (if not completely) responsible for the scatter in the $T - \sigma$ relation.

Struble & Ftaclas (1994) listed the apparent ellipticities, as obtained by optical analysis, for a very large cluster sample. In the sample we studied there is no significant correlation between these ellipticities and the (absolute) residuals of our $\sigma$-$T$ or the values of $\beta_{spec}$. However, this correlation could be easily concealed by the indirect correspondence between real and apparent ellipticity, by the presence of substructures and projection effects, as well as by measurement errors.

### 3.7. PHYSICAL INTRINSIC DISPERSION

The (possible) intrinsic dispersion in the $\sigma$-$T$ relation could be due to real physical differences among different clusters, i.e. different values of $\beta_{spec}$ pertaining to clusters with different properties.

We found no significant correlation between (absolute) residuals from the relation (or $\beta_{spec}$ values) and several cluster properties: richness class, Bautz-Morgan type and Rood-Sastry type (Abell, Corwin, & Olowin 1989, Struble & Rood 1987, Struble & Ftaclas 1994), and mass flow rate in cooling flows (Stewart et al. 1984; Edge, Stewart, & Fabian 1992; White private communication 1994). However, some possible physical correlations might be concealed by the effect of cluster asphericity (see § 3.6) or other spurious effects.

The matter could be clarified by the fact that the galaxy/gas energy ratio is predicted to be the same both when obtained from spectral data, $\beta_{spec}$, and when obtained from fitting geometrical cluster quantities, $\beta_{fit}$, on the assumption that the galaxies and the gas are in hydrostatic equilibrium within the cluster potential. However, the precise formulation of $\beta_{fit}$ depends on additional hypotheses (see Gerbal, Durret, & Lachiéze-Rey 1994 and references therein). It is claimed that the discrepancy between $\beta_{spec}$ and $\beta_{fit}$ (e.g. Edge and Stewart 1991a) disappears when less restrictive hypotheses are considered (Bahcall & Lubin 1994 and Gerbal, Durret, & Lachiéze-Rey 1994).

Our average value of $\beta_{spec} = 1.03$ is still consistent with the value of $\beta_{fit} = 0.84 \pm 0.1$ proposed by Bahcall & Lubin (1994). However, we suggest a further re-formulation of $\beta_{fit}$ for a fruitful comparison between average or individual $\beta$ values. In fact, previous works assumed isothermal galaxy distribution ($d\sigma_r^2/dr = 0$, where $\sigma_r$ is the radial component of the spatial velocity dispersion) and often also null anisotropy, but these hypotheses are no longer valid in the presence of strong velocity anisotropy as (possibly) suggested by our cluster VDPs.

### 4. SUMMARY AND CONCLUSIONS

We analyzed a cluster sample characterized by the homogeneity of the redshift data in each cluster. We used robust estimates for location and scale in the velocity distribution and a nonparametric method for the selection of cluster members in redshift space. Actually, if the swift increase/decrease of observed VDP in the central regions of some clusters is an indication of strong velocity anisotropies, we cannot expect the velocity distribution to be gaussian. So the choice of non parametric methods seems the most reasonable one.

We present the 2-D optical maps of our sample clusters, obtained by the kernel adaptive technique, using an optimized smoothing parameter.

The clusters with the strong presence of substructures, as shown by a bimodal distribution of galaxy redshifts, were rejected from the final sample. Less apparent substructures were parametrized by applying two different tests.

The value of the integrated $\sigma$ is reliable if computed on a magnitude complete sample; otherwise, a different sampling in different cluster regions may bias the observed value of $\sigma$ towards the local value corresponding to the region with deeper sampling. For our clusters with available galaxy magnitudes, we extracted subsamples complete in magnitude.

Our analysis lends support to the scenario in which late-type galaxies may not be in dynamical equilibrium with the cluster, since 8 of the 17 clusters we analyzed show significant differences in mean- and/or $\sigma$-values between early- and late-type galaxies. For the 5 clusters, which show significant difference in $\sigma$, the $\sigma$ value computed using the global population is larger by about 190 km/s, on average, with respect to the $\sigma$ value computed using the early-galaxies. Moreover, the effect concerns poorer and cooler clusters. In fact, we found kinematical differences for 6 of 10 clusters with $kT \leq 3.9$ keV (the median value of $kT$ in the sample), but only for 2 of 7 clusters with $kT > 3.9$ keV. One can hypothesize that in the poorer and cooler clusters the spirals are still infalling, but in the richer and hotter clusters the spirals have already reached virial equilibrium. Neglecting this effect could strongly affects the slope of the $\sigma$ - $T$ relation.

Our analysis of integral VDP shows that the value of $\sigma$ is dependent on the radius at which it is computed. The average of the absolute differences between $\sigma$ as computed at 0.5 $h^{-1}$ $Mpc$ and the total value of $\sigma$ is $\sim$ 90 km s$^{-1}$ ; for a few clusters this difference is larger than 300 km s$^{-1}$ . The increase/decrease of VDP in internal cluster regions suggests the possible presence of velocity anisotropies.

We found that cluster VDP can be strongly affected by the influence of close clusters (as in the case of A3391/A3395, see § 3.3). The presence of substructures, filaments and superclusters does not seem to induce an asymmetric velocity field in the cluster. This can be checked by noting that significant velocity gradients are found only in 4 out of 38 clusters.

As a consequence of our accurate analysis of velocity dispersion, the scatter in the $\log\sigma$- $\log T$ plane is reduced by more than 30% with respect to previous determinations. The gas/galaxy energy equipartition model, $\beta_{spec} = 1$, is consistent with our data; this confirms previous results (Edge & Stewart 1991a; Lubin & Bahcall 1993), but is in disagreement with Bird, Mushotzky, & Metzler (1995). We obtain an average value of $\beta_{spec} = 1.03 \pm 0.05$.

The scatter in the $\sigma$-$T$ relation suggests that part of the dispersion is intrinsic. This intrinsic dispersion may have several different origins. Part of the scatter can be explained by the effect of cluster asphericity, while part may be physical, different values of $\beta_{spec}$ pertaining to clusters with different properties. Yet, in our sample no significant correlations are found between the (absolute) residuals of the fitted $\sigma - T$ relation and the projected ellipticity, substructure parameters, richness class, Bautz-Morgan type and Rood-Sastry type, and mass flow rate in cooling flows.

We are particularly indebted to Armando Pisani for enlightening discussions on the method of adaptive kernels and the determination of density peaks. We thank José M. Solanes for having mailed us an electronic copy of his cluster data; Cathy Clemens and John Huchra for having kindly ftp-ed us the recent version of the zcat catalog (Huchra et al. 1992); David A. White for having provided us with his still-unpublished data; and the referee, Alastair Edge, for interesting suggestions.

This work was partially supported by the *Ministero per l'Università e per la Ricerca scientifica e tecnologica*, and by the *Consiglio Nazionale delle Ricerche (CNR-GNA)*.

## APPENDICES

## A ADAPTIVE KERNEL DENSITY

If we have a sample of data $\vec{\xi}_1, \vec{\xi}_2, \ldots, \vec{\xi}_N$ in $d$ dimensions, we can define an empirical distribution function:

$$g(\vec{\xi}) = \frac{1}{N} \sum_{i=1,N} \delta(\vec{\xi} - \vec{\xi}_i). \quad (4)$$

It contains all the information in the sample but is not a satisfactory estimate of the true distribution $f$. In fact, the short wavelength behavior of $g(\vec{\xi})$ is a consequence of discrete sampling. A smooth estimate $\hat{f}$ may be obtained by convolving $g$ with a probability density function $K(\vec{x}|\vec{\xi})$, kernel of the integral function:

$$\hat{f}(\vec{x}) = \int_{R^d} g(\vec{\xi}) K(\vec{x}|\vec{\xi}) d\vec{\xi} = \frac{1}{N} \sum_{i=1,N} K(\vec{x}|\vec{\xi}_i), \quad (5)$$

where $R^d$ is the d-dimension Real region of integration. As kernel we chose a function $K(\vec{x}|\vec{\xi}_i) = K(\vec{x}; \mu = \vec{\xi}_i, \sigma)$ with $\mu = \vec{\xi}_i$ (position of the datum) and the smoothing parameter $\sigma$ large enough for $\hat{f}$ to be a smooth function. We obtained an estimate of $f$ by eliminating the short-wavelength behavior of $g$ without modifying its long-wavelenghts behavior. Obviously too large a value of $\sigma$ will oversmooth $\hat{f}$, hiding its true features. There seems to be general agreement that the choice of the kernel shape is not important; hence we decided to use a gaussian, because of its analytical properties (see, e.g., Merritt & Tremblay, 1994). The choice of the value of the smoothing parameter is crucial. A single value of $\sigma$ will be too large to describe fairly a rapidly changing true probability density, and too small where the probability density is smoother. To avoid this problem several authors have introduced an adaptive kernel function, with the smoothing parameter $\sigma$ sensitive to the local density of data. Silvermann (1986), once he obtained a pilot estimate $f_p$ with a fixed $\sigma$, defines:

$$\sigma_i = \left( \frac{\left[\prod_{j=1,N} f_p(\vec{x}_j)\right]^{\frac{1}{N}}}{f_p(\vec{x}_i)} \right)^{\alpha} \sigma. \quad (6)$$

Hence the greater $f_p(\vec{x}_i)$, the lower $\sigma_i$. The sensitivity parameter $\alpha$ is usually fixed to 1/2.

The problem of choosing an optimal and objective value of $\sigma$ still remains. Pisani (1993), following an idea developed by Stone (1984), proposes minimizing the integrated square error:

$$ISE = \int_{R^d} (f(\vec{x}) - \hat{f}(\vec{\xi}))^2 d\vec{x}, \quad (7)$$

which is equivalent to minimizing the quantity:

$$M_f(\sigma) = ISE - \int_{R^d} f^2 d\vec{x} = \int_{R^d} \hat{f}^2 d\vec{x} - 2\int_{R^d} f\hat{f}d\vec{x}. \quad (8)$$

The quantity $f\hat{f}$ is unknown, but we can note (Stone 1984) that:

$$\int_{R^d} f\hat{f}d\vec{x} = \frac{1}{N}\sum_{i=1,N}\int_{R^d} K(\vec{x}-\vec{x}_i)f(\vec{x})d\vec{x}. \quad (9)$$

The expectation value of this quantity is the expectation of the average of $\int_{R^d} K(\vec{x}-\vec{y})f(\vec{x})d\vec{x}$. But the expectation of the average of a variable is equal to the mean of the same variable:

$$E\left[\frac{1}{N}\sum_{i=1,N}\int_{R^d} K(\vec{x}-\vec{x}_i)f(\vec{x})d\vec{x}\right] = \int_{R^d} f(\vec{y})d\vec{y}\int_{R^d} K(\vec{x}-\vec{y})f(\vec{x})d\vec{x}. \quad (10)$$

And, reversing the same argument with the probability function $f(\vec{x})f(\vec{y})$:

$$E\left[\int_{R^d} f\hat{f}d\vec{x}\right] = E[K(\vec{x}-\vec{y})] = E\left[\frac{1}{N(N-1)}\sum\sum_{i\neq j} K(\vec{x}_i-\vec{y}_j)\right], \quad (11)$$

where $\vec{x}$ and $\vec{y}$ are independent variables each having density $f$. That leads to the unbiased estimate:

$$\int_{R^d} f\hat{f}d\vec{x} = \frac{1}{N(N-1)}\sum\sum_{i\neq j} K(\vec{x}_i-\vec{y}_j). \quad (12)$$

Hence we are able to express $M_f(\sigma)$ simply by using the known estimate $\hat{f}$. The function $M_f(\sigma)$ has a minimum and the corresponding $\sigma$-value is the optimal smoothing parameter.

## A1. SOME QUANTITIES USED IN THE PAPER

Having evaluated the density, following the method of clustering analysis introduced by Pisani (1993) and its multivariate extension (Pisani 1995), we are able to:

- associate each object with its own density peak;
- calculate the significance of a peak;
- compute the probability that an object could belong to a peak, and hence the overlapping between two peaks;
- generate a sequence of objects with decreasing density starting from a peak top.

It is possible to associate a galaxy with a peak by using the sequence:

$$\vec{x}_{i+1} = \vec{x}_i + \left[\frac{d}{\sum_{j=1}^{N}\left(\frac{\vec{\nabla}\hat{f}(\vec{x}_j)}{\hat{f}(\vec{x}_j)}\right)^2}\right]\frac{\vec{\nabla}\hat{f}(\vec{x}_i)}{\hat{f}(\vec{x}_i)}, \quad (13)$$

where $d$ is the number of dimensions. Usually we obtain isolated objects, which permit us to define $\sigma_0 = \max\{\sigma_i\}$ in the field. We can see the density $f(\vec{x})$ as the realization of $\nu$ different peaks:

$$f(\vec{x}) = \sum_{\mu=0,\nu} f_\mu(\vec{x}) \quad (14)$$

where $f_\mu(\vec{x}) = \frac{1}{N}\sum_{i\in\mu} K(\vec{x};\vec{x}_i,\sigma_i)$ is the probability density of the $\mu$-th peak and $f_0(\vec{x}) = \frac{1}{N}K(\vec{x};\vec{x}_i,\sigma_0)$ is the probability density of the i-th isolated object. We can define the likelihood:

$$L_N = \prod_{i=1,N} f(\vec{x}_i) \quad (15)$$

and compute the significance of the $\mu$-th peak by calculating:

$$\chi^2 = -2\ln\left(\frac{L(\mu)}{L_N}\right), \quad (16)$$

where $L(\mu)$ is the value that $L_N$ would have if each object of the $\mu$-th peak were described by $f_0(\vec{x})$, i.e. if it belonged to the field (Materne 1979).

The probability that an object belongs to the field can be evaluated by:

$$P(i \in 0) = \frac{\frac{1}{N}K(\vec{x};\vec{x}_i,\sigma_0)}{f(\vec{x}_i)}, \quad (17)$$

and, similarly, the probability of belonging to the $\mu$-th peak is:

$$P(i \in \mu) = \delta\frac{f_\mu(\vec{x}_i)}{f(\vec{x}_i)}, \quad (18)$$

where $\delta$ is equal to $1 - P(i \in 0)$ from the normalization condition $\sum_\mu P(i \in \mu) + P(i \in 0) = 1$. We can thus evaluate the overlapping between two peaks $\mu,\nu$ counting the objects associated with peak $\mu$ and with $P(i \in \nu) > P(i \in 0)$ and the objects of $\nu$ with $P(i \in \mu) > P(i \in 0)$.

To obtain a sequence of objects with decreasing density, we start from the object closest to the top of a peak (Kittler 1976 and Pisani 1995). Then we define as neighbours all the objects at $\vec{x}_i$ closer to the starting point at $\vec{x}_1$ than the sum $\sigma_1 + \sigma_i$. We choose as our second point the object with highest density among the neighbours. Then we add to the set of neighbours all those of the second point and iterate the procedure. When we find, as a term of the sequence, an object associated with another peak, we restart the procedure from this peak. Then, when we have as a sequence term another object of the previous peak, we merge the two peaks and continue the procedure.

This way it is possible to study the VDP by taking into account the ellipticity of a cluster or to analyse the VDP of a structured cluster by distinguishing the presence of substructures (see also Pisani 1995).

## B PROJECTION FACTOR

We follow the formalism and the content of paragraphs 2.5 and 4.3 of Binney and Tremaine (1987, BT). We assume that clusters are axisymmetric (prolate or oblate), that isodensity surfaces are concentric, similar ellipsoids, and that galaxies are distributed in a way similar to the total binding mass. We also assume that the virial theorem holds, that rotation is negligible (see, e.g., Rood et al. 1972; Gregory & Tifft 1976; Dressler 1981), and that ellipticity is produced by velocity anisotropy and not by tidal interaction (as, on the contrary, proposed by Salvador-Solé & Solanes 1993). If, as suggested by these authors, the ellipticity were produced by a tidal effect, but the kinetic energy tensor is isotropic, there would be no dependence of the velocity dispersion on the direction of the line of sight. Since, on the contrary, we want to estimate an upper limit to the sensitivity of the velocity dispersion to the position of the observer, we neglect tidal elongation and assume that the shape of clusters is entirely due to anisotropy in the kinetic energy tensor.

We use the tensor virial theorem to link the shape of the cluster to its internal motion. We define an orthogonal reference system $x_1$, $x_2$, $x_3$, coincident with the principal axes of the ellipsoid, $x_3$ being the symmetry axis of the system. If the kinetic energy of galaxies is due to their random motions, the kinetic energy tensor is given by $\frac{1}{2}\Pi_{jk}$ (see eq. 4.74 of BT), and the tensor virial theorem is reduced to $\Pi_{jk} + W_{jk} = 0$, where $W_{jk}$ is the potential energy tensor (see eq. 2-123 of BT). From the symmetry assumed, we have $\Pi_{11} = \Pi_{22} = M\sigma_0^2$, where $M$ is the total mass of galaxies and $\sigma_0$ is the velocity dispersion along a line of sight lying in the equatorial plane of the cluster. The component along the symmetry axis is written as $\Pi_{33} = (1 + \delta)\Pi_{11}$. The total kinetic energy of the system is $K = \frac{1}{2}\sum_j(\Pi_{jj}) = \frac{1}{2}M\sigma_{3D}^2$, so that one obtains $\sigma_{3D} = \sigma_0\sqrt{3 + \delta}$. If the line of sight forms an angle $i$ with the $x_3$ axis, the observed velocity dispersion $\sigma_{los}$ is linked to $\sigma_0$ by the relation $\sigma_{los}^2 = \sigma_0^2 \sin^2 i + \sigma_0^2(1 + \delta)\cos^2 i$ (compare eq. 4.96b of BT, but notice the different assumption on the sign of $\delta$). Combining this with the above relations gives

$$\sigma_{3D} = \sigma_{los}\sqrt{\frac{3 + \delta}{1 + \delta \cos^2 i}} \qquad (19)$$

Now, from the tensor virial theorem and from the above-mentioned relation, it turns out that

$$1 + \delta = \frac{\Pi_{33}}{\Pi_{11}} = \frac{W_{33}}{W_{11}} = \frac{a_3^2 A_3}{a_1^2 A_1}, \qquad (20)$$

where the $a_i$ are the ellipsoid semi-axes and the $A_i$ are the expressions given in Table 2.1 of BT. This result, due to the assumptions we made, is entirely independent of the radial density distribution in the cluster (see eq. 2.134 of BT and the following discussion).

If the cluster is prolate ($a_1 = a_2 < a_3$), with the axial ratio $\beta = a_1/a_3$, Table 2.1 of BT gives

$$\delta = \frac{2}{\beta^2}\frac{\ln\left(\frac{\beta}{1-\sqrt{1-\beta^2}}\right) - \sqrt{1-\beta^2}}{\frac{\sqrt{1-\beta^2}}{\beta^2} - \ln\left(\frac{\beta}{1-\sqrt{1-\beta^2}}\right)} - 1, \qquad (21)$$

so that the projection factor for the velocity dispersion $f_p = \sigma_{3D}/\sigma_{los}$, using equations (19) and (21), may be expressed as a function of the intrinsic axial ratio and the inclination of the cluster.

In a similar way, if the cluster is oblate ($a_1 = a_2 > a_3$), with the axial ratio $\beta = a_3/a_1$, one obtains

$$\delta = 2\beta^2 \frac{\frac{1}{\beta} - \frac{\arcsin\sqrt{1-\beta^2}}{\sqrt{1-\beta^2}}}{\frac{\arcsin\sqrt{1-\beta^2}}{\sqrt{1-\beta^2}} - \beta} - 1. \qquad (22)$$

The value of $\delta$ is positive for prolate clusters and negative for oblate ones.

## NOTES TO THE TABLES

### Table 1: The Initial Cluster Sample

**References:** [1] Malumuth et al. (1992); [2] Beers et al. (1991); [3] Fabricant et al. (1993); [4] Hill, & Oegerle (1993); [5] Giovannelli et al. (1982); [6] Gregory, & Thompson (1978); [7] Moss, & Dickens (1977); [8] Beers et al. (1992); [9] Kent, & Sargent (1983); [10] Quintana, & Ramirez (1990); [11] Ostriker et al. (1988); [12] Dressler, & Shectman (1988b); [13] Hintzen et al. (1982); [14] Richter (1987); [15] Richter (1989); [16] Geller et al. (1984); [17] Gavazzi (1987); [18] Tifft (1978); [19] Dickens, & Moss (1976); [20] Kent, & Gunn (1982); [21] Teague et al. (1990); [22] Quintana et al. (1985); [23] Oegerle, & Hill (1993) [24] Gregory, & Thompson (1984); [25] Fabricant, Kent, & Kurtz (1989); [26] Hintzen (1980); [27] Pinkney et al. (1993) [28] Zabludoff, Huchra, & Geller (1990); [29] Bothun, & Schombert (1988); [30] Butcher, & Olmer(1985); [31] Scodeggio et al.(1994); [32] Sharples, Ellis, & Gray (1988); [33] Chincarini, Tarenghi, & Bettis (1981); [34] Lucey et al. (1983); [35] Lauberts, & Valentjin, E.A. (1989); [36] Dickens, Currie, & Lucey (1986); [37] Metcalfe, Godwin, & Spenser (1987); [38] Bardelli et al. (1994); [39] Quintana & de Souza (1993); [40] Sodrè et al. (1992); [41] Malumuth, & Kriss (1986); [42] Beers et al. (1984); [43] Binggeli, Sandage & Tammann (1985); [44] Elvis et al. (1992); [45] Abramopoulos, & Ku (1983); [46] Edge, & Stewart (1991b); [47] HEASARC Archive; [48] Beers, & Tonry (1985); [49] Breen et al. (1994); [50] McMillan, Kowalski & Ulmer (1989); [51] Rhee, & Latour (1991); [52] Miyaji et al. (1993); [53] Pierre et al. (1994); [54] Piro, & Fusco-Femiano (1988).

### Table 2: Effects of Galaxy Morphology

**References:** [1] Fabricant et al. (1993); [2] Butcher & Oemler (1985); [3] Dressler (1980); [4] Huchra et al. (1992); [5] Paturel et al. (1989); [3] Dressler (1980); [6] Biviano (1986); [7] Poulain, Nieto, & Davoust (1992); [8] Ostriker et al. (1988); [9] Dressler, & Schectman (1988b); [10] Richter (1987); [11] Richter (1989); [12] Kent, & Gunn (1982); [13] Sharples, Ellis, & Gray (1988); [14] Bell, & Whitmore (1989); [15] Binggeli, Sandage & Tammann (1985).